\documentclass[3p,times,procedia]{elsarticle}
\usepackage{nupha_ecrc,amsmath}
\usepackage{wrapfig}

\volume{00}

\firstpage{1}

\journalname{Nuclear Physics A}

\runauth{}

\jid{nupha}

\jnltitlelogo{Nuclear Physics A}

\usepackage{amssymb}

\usepackage[figuresright]{rotating}

\newcommand{\ket}[1]{|#1\rangle}

\newcommand{\di}{{\rm d}}
\newcommand{\ii}{i}

\def\spt{{\cal S}}
\def\wT{{\widehat T}}
\def\wj{{\widehat j}}
\def\wspt{{\widehat{\cal S}}}

\def\wPhi{{\widehat{\Phi}}}

\def\wpsi{{\widehat{\psi}}}
\def\wrho{{\widehat{\rho}}}

\def\wP{{\widehat{P}}}
\def\wJ{{\widehat{J}}}

\newcommand{\tr}{{\rm tr}}

\newcommand{\x}{{\rm x}}

\newcommand{\be}{\begin{equation}}
\newcommand{\ee}{\end{equation}}                                                                               
\newcommand{\bea}{\begin{eqnarray}}
\newcommand{\eea}{\end{eqnarray}}

\begin{document}

\begin{frontmatter}



\dochead{XXVIIIth International Conference on Ultrarelativistic Nucleus-Nucleus Collisions\\ (Quark Matter 2019)}

\title{Does the spin tensor play any role in non-gravitational physics?}


\author{F. Becattini}

\address{Department of Physics and Astronomy of the University of Florence,
Via G. Sansone 1, I-50019, Sesto Fiorentino (Firenze), Italy}

\begin{abstract}
In view of the recent polarization measurements in ultra-relativistic heavy ion 
collisions, we discuss the possibility of a physical meaning of the spin angular 
momentum in quantum field theory and relativistic hydrodynamics. 
\end{abstract}




\end{frontmatter}


\section{Introduction}
\label{intro}

The evidence of a finite global polarization of particles in relativistic heavy-ion 
collisions \cite{STAR} has opened a new perspective in the field as well as in the 
theory of relativistic matter. While the experiments proved to be able to measure 
polarization differentially in momentum space \cite{STAR2},
theoretical predictions are mostly based on local thermodynamic equilibrium, which
imply a relation between spin and thermal vorticity \cite{becaspin}.
In fact, this relation has been recently questioned based on the idea that the 
separation between orbital and spin angular momentum in quantum physics is possibly
physically meaningful. In this case, the spin of the particles would not be necessarily 
related to thermal vorticity \cite{florka,florkb}.  
In gravitational physics, the stress-energy tensor and the spin tensor (Einstein-Cartan 
theory) have an objective physical meaning, because they are related to the local
geometry of space-time. Beyond gravitational theories, the 
problem of the physical significance of the spin tensor beyond is a long-standing 
one \cite{hehl} and has been rediscussed more recently in e.g. ref.~\cite{becatinti2}, 
where it was demonstrated that, e.g. that some quantities such as viscosity change
by small quantum terms depending on whether a spin tensor is there. 
In this work, which is largely based on ref.~\cite{becaflork}, I will delve into 
this subject and the possible phenomenological implications.

\section{Spin tensor and pseudo-gauge transformations}
\label{spint}

In relativistic quantum field theory in flat space-time, according to Noether's theorem, for 
each continuous symmetry of the action, there is a corresponding conserved current. 
The currents associated with the translational symmetry and the Lorentz symmetry
are the so-called {\em canonical} stress-energy tensor and the {\em canonical} angular 
momentum tensor:
\be\label{cantens}
  \wT^{\mu\nu}_C = \sum_a \frac{\partial \cal L}{\partial(\partial_\mu \wpsi^a)} \partial^\nu \wpsi^a
   - g^{\mu\nu} {\cal L}, \qquad \qquad
  \widehat{\cal J}^{\mu,\lambda\nu}_C = x^\lambda \wT^{\mu\nu}_C - x^\nu \wT^{\mu\lambda}_C 
   + \wspt^{\mu,\lambda\nu}_C.
\ee
In eq.~\eqref{cantens}, ${\cal L}$ is the lagrangian density, while $\wspt_C$ reads
\begin{eqnarray} \label{SC}
   \wspt^{\mu,\lambda\nu}_C =  - \ii \sum_{a,b} \frac{\partial \cal L}{\partial(\partial_\mu \wpsi^a)} 
    D(J^{\lambda\nu})^a_b \wpsi^b
\end{eqnarray}  
with $D$ being the irreducible representation matrix of the Lorentz group pertaining 
to the field. The above tensors fulfill the following equations:
\be\label{conserv}
\nabla_\mu \wT^{\mu\nu}_C = 0, 
\qquad \nabla_\mu  \widehat{\cal J}^{\mu,\lambda\nu}_C
= \wT^{\lambda\nu}_C - \wT^{\nu\lambda}_C + \nabla_\mu \wspt^{\mu,\lambda\nu}_C = 0,
\ee 

It turns out, however, that the stress-energy and angular momentum tensors are not 
uniquely defined. Different pairs can be generated either by just changing the lagrangian 
density or, more generally, by means of the so-called pseudo-gauge transformations \cite{hehl}:
\begin{eqnarray}\label{transfq}
 && \wT^{\prime \mu \nu} = \wT^{\mu \nu} +\frac{1}{2} \nabla_\lambda
 \left( \wPhi^{\lambda, \mu \nu } - \wPhi^{\mu, \lambda \nu} - 
 \wPhi^{\nu, \lambda \mu}  \right), \nonumber \\
 && \wspt^{\prime \lambda, \mu \nu} = \wspt^{\lambda,\mu\nu}-\wPhi^{\lambda,\mu\nu},
\end{eqnarray}
where $\wPhi$ is a rank-three tensor field antisymmetric in the last two indices,
often called and henceforth referred to as {\em superpotential}. In Minkowski 
space-time, the newly defined tensors preserve the total energy, momentum, and angular 
momentum (herein expressed in Cartesian coordinates):
\begin{eqnarray}\label{total}
  \wP^\nu = \int_\Sigma \di \Sigma_\mu \wT^{\mu\nu}, \qquad 
 \wJ^{\lambda\nu} = \int_\Sigma \di \Sigma_\mu \widehat{\cal J}^{\mu,\lambda\nu},
\end{eqnarray}
as well as the conservation equations (\ref{conserv}) \footnote{This statement only
applies to Minkowski space-time, in generally curved space-times it is no longer
true \cite{hehl}.}.

A special pseudo-gauge transformation is the one where one starts with the canonical 
definitions and the superpotential is the spin tensor itself, that is, $\wPhi = \wspt$. 
In this case, the new spin tensor vanishes, $\wspt^\prime = 0$, and the new stress-energy tensor 
is the so-called Belinfante stress-energy tensor $\wT_B$,
\be\label{belinf}
 \wT^{\mu \nu}_B = \wT^{\mu \nu}_C +\frac{1}{2} \nabla_\lambda
 \left( \wspt^{\lambda, \mu \nu}_C - \wspt^{\mu, \lambda \nu}_C - 
 \wspt^{\nu, \lambda \mu}_C  \right).
\ee

It is common wisdom in Quantum Field Theory that no actual physical measurement can be
made in flat space-time discriminating between different couples of stress-energy 
and spin tensor; in other words, no measurement can depend on the superpotential. 
This is effectively rephrased in the adage {\em it is impossible to
separate orbital angular momentum and spin}. One should be a little more specific:
when saying "no actual physical measurement" direct measurements of the energy-momentum
{\em density}, which are obviously sensitive to the superpotential according to 
eq.~\eqref{transfq}, are not included. Indeed, space-time local measurement is not possible 
in a high energy physics experiment and in heavy ion collisions as well. Upon some
reflection, one can realized that any measurement involves particles in momentum 
space and momentum spectra are supposedly independent of the pseudo-gauge transformation.

We can reinforce the above argument by saying that the stress-energy tensor, being
a pseudo-gauge dependent quantity, is not a physical object; it is defined, and can
be probed with particle measurements, only up to quantum corrections encoded in
the superpotential $\wPhi$. This might be the end of the story, however we are going 
to see that something more can be told.

\section{States, operators and local equilibrium}
\label{stopleq}

In quantum physics, there are operators and states. Operators can be invariant
under some transformation; for instance, the operators in \eqref{total} are invariant
under the transformation \eqref{transfq}. On the other hand, quantum states can 
either be invariant under the same transformation or they can not. This 
dichotomy is reminiscent of the general definition of spontaneous symmetry breaking: 
while the action of a quantum field theory is invariant under some transformation 
of the fields, the vacuum state is not. Similarly, within our scope, we can conceive a 
quantum states which are {\em not} invariant under a pseudo-gauge transformations. 
In principle, this is a very easy operation; take a free field theory, whose Hilbert 
space basis states are defined by a set of number of occupations for each momentum 
$p$ and spin state $s$:
$$
 \ket{ \{n\}_{p,s} }
$$
These states, being eigenstates of the four-momentum, are clearly pseudo-gauge invariant.
Nevertheless, a pseudo-gauge-dependent state can be generated by just forming a 
linear superposition of the above state with complex coefficients which functionally 
depend on a particular superpotential $\Phi$ in eq.~\eqref{transfq}, once a reference
set is chosen (e.g. Belinfante):
\be\label{state}
  \ket{ {\rm state} } = \sum C[\Phi]_{ \{n\}_{p,s} } \ket{ \{n\}_{p,s} }
\ee
This can be easily extended to a mixed state and its associated density operator $\wrho$. 
Hence, if the density operator depends on the superpotential, so will the mean values
of {\em any} quantum operator $\widehat O$, whether it is related to measurable 
quantity or not:
$$
 O([\Phi]) = \tr \left( \wrho[\Phi] \, \widehat O \right)
$$

To make this concrete, we will now write down a mixed state which is very relevant
for relativistic fluid, which is indeed pseudo-gauge dependent. This is the density 
operator describing local thermodynamic equilibrium in quantum field theory 
\cite{zubarev,weert,betaframe,hongo}:
\be\label{leqd}
\wrho_{\rm LE} = \frac{1}{Z} \exp \left[-\int_\Sigma \di \Sigma_\mu \left(\wT^{\mu\nu}_B \beta_\nu 
 - \zeta \, \wj^\mu \right) \right],
\ee
where $\beta$ and $\zeta$ are the four-temperature vector and the ratio between local 
chemical potential and temperature respectively. The operator \eqref{leqd} is obtained 
by maximizing the entropy with the constraints \cite{betaframe}:
\be\label{constr}
n_\mu \tr \left(\wrho \, \wT_B^{\mu\nu}\right) = n_\mu T_B^{\mu\nu}, 
\qquad n_\mu \tr \left(\wrho \, \wj^{\mu}\right) = n_\mu j^{\mu}.
\ee
Note the use of the Belinfante's stress-energy tensor $\wT_B$ in both eqs.~\eqref{leqd}
and \eqref{constr}; in this case, angular momentum density constraints are redundant 
because the associated spin tensor vanishes \cite{becaflork}. Note that the 
operator (\ref{leqd}) is not the actual density operator, because it is not generally 
stationary as required in the Heisenberg picture. In fact, the true density operator 
for a system achieving local thermodynamic equilibrium is the one in eq.~\eqref{leqd}
for some 3D hypersurface at the time $\tau_0$, that is $\Sigma(\tau_0)$ \cite{zubarev}
(see also discussion in ref.~\cite{becazuba}).

One can now use the pseudo-gauge transformations of Sec.~\ref{spint} to rewrite
the local thermodynamic equilibrium density operator as a function of, e.g., canonical 
tensors. Using eq.~(\ref{belinf}), 
\be\label{leqdC2}
\wrho_{\rm LE} = \frac{1}{Z} \exp \left[-\int_\Sigma \di \Sigma_\mu \left(\wT^{\mu\nu}_C 
  \beta_\nu - \frac{1}{2} \varpi_{\lambda\nu} \wspt^{\mu, \lambda \nu}_C - \frac{1}{2}
  \xi_{\lambda\nu} \left( \wspt^{\lambda, \mu \nu}_C + \wspt^{\nu, \mu \lambda}_C \right)  
   - \zeta \wj^\mu\right) \right],
\ee
where
\be\label{thvort}
\varpi_{\lambda\nu} = \frac{1}{2} (\nabla_\nu \beta_\lambda - \nabla_\lambda \beta_\nu)
\qquad \qquad
\xi_{\lambda\nu} = \frac{1}{2} (\nabla_\nu \beta_\lambda + \nabla_\lambda \beta_\nu)  
\ee
are the thermal vorticity and the symmetric part of the gradient of the four-temperature 
vector, what can be called {\em thermal shear tensor}. 

Had we used another couple of tensors, instead of Belinfante, to calculate the
local thermodynamic equilibrium operator, the operator would be different. The   
inclusion of angular momentum density amongst the constraints defining local thermodynamic 
equilibrium implies an additional constraint:
\be\label{constr3}
   n_\mu \tr \left(\wrho \, \wspt^{\mu,\lambda\nu}\right) = n_\mu \spt^{\mu,\lambda\nu}.
\ee
and the introduction of an additional antisymmetric tensor field $\Omega_{\lambda\nu}$ 
as Lagrange multiplier for the equation \eqref{constr3}, the {\em spin potential}. 
The associated local equilibrium density operator is:
\be\label{leqd-s}
\wrho_{\rm LE} = \frac{1}{Z} \exp \left[-\int_\Sigma \di \Sigma_\mu \left(\wT^{\mu\nu} \beta_\nu 
 - \frac{1}{2} \Omega_{\lambda\nu} \wspt^{\mu,\lambda\nu} - \zeta \wj^\mu\right) \right].
\ee
If $\wT=\wT_C$ and $\wspt=\wspt_C$, 5he operator \eqref{leqd-s} is the same as 
\eqref{leqdC2} (hence \eqref{leqd}) if $\beta$ and $\zeta$ are the same fields, 
$\Omega=\varpi$ and $\xi=0$. The latter condition implies that the equilibrium 
should be global \cite{becacov} for the two operators to coincide. Otherwise, they 
do not; the operator is pseudo-gauge dependent and we are in a situation like the 
one described by the eq.~\eqref{state} for a pure state. 

\begin{wrapfigure}{r}{0.5\textwidth}
\includegraphics[keepaspectratio, width=0.4\columnwidth]{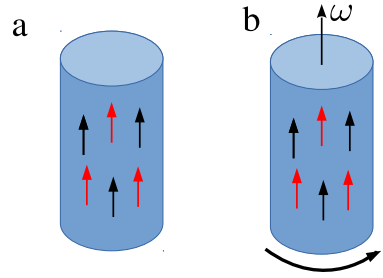}
  \caption{(Color online) a) a macroscopic fluid at rest with polarized particles 
  (black arrows) and antiparticles (red arrows) b) if the spin tensor vanishes, the fluid 
  must be rotating (non-vanishing thermal vorticity) to have polarized particles and 
  antiparticles in the same direction.}
  \label{figura}
\end{wrapfigure}

To get an insight of which state an operator \eqref{leqd-s} can describe that 
cannot be described by the eq.~\eqref{leqd}, one can envisage a fluid temporarily at rest 
with a constant temperature $T$, hence $\beta = (1/T) (1,{\bf 0})$, wherein both 
particles and antiparticles are polarized in the same direction (see Fig.~\ref{figura}). 
Such a situation cannot be described as local thermodynamic equilibrium by the density 
operator (\ref{leqd}) because the only way to get particles and antiparticles polarized, 
in this case, is through a non-vanishing thermal vorticity, that is when the fluid
rotates. Yet, thermal vorticity vanishes if $\beta$ is constant, so only through 
a non-vanishing spin potential $\Omega$ is a charge-independent polarization (such 
as that observed in the experiment \cite{STAR}) possible. 

Indeed, polarization is the most sensitive variable to discriminate between the
two local equilibria density operators. It is thus very important to understand
whether the data can be accomodated within the Belinfante LEQ description \eqref{leqd}
or whether an extension of hydrodynamics to include the spin potential is needed.
The latter option is under scrutiny \cite{flork2}. For an ampler discussion
we refer the reader to ref.~\cite{becaflork}.

\section*{Acknowledgments}

Very useful discussions with G. Q. Cao, W. Florkowski, K. Fukushima, E. Grossi, 
X. G. Huang, E. Speranza are gratefully acknowledged.


\end{document}